\documentclass[twocolumn,showpacs,preprintnumbers]{revtex4}
\usepackage{amsmath}
\usepackage{graphicx}
\usepackage{dcolumn}
\usepackage{bm}

\input{tcilatex}
\begin{document}

\preprint{}
\title{Breakdown of Ehrenfest theorem for free particle constrained on a
hypersurface}
\author{Q. Li, Z. Li, X. Wang,}
\author{Q. H. Liu}
\email{quanhuiliu@gmail.com}
\affiliation{School for Theoretical Physics, School of Physics and Electronics, Hunan
University, Changsha 410082,China}
\date{\today }

\begin{abstract}
There is a belief that the Ehrenfest theorem holds true universally. We
demonstrate that for a classically nonrelativistic particle constrained on
an $N-1$ ($N\geq 2$) curved hypersurface embedded in $N$ flat space, the
theorem breaks down.
\end{abstract}

\maketitle

For a free nonrelativistic particle constrained on an ($N-1$)-dimensional
smooth curved surface $\Sigma ^{N-1}$ in flat space $R^{N}$ ($N\geq 2$), in
classical mechanics we have two well-defined forms of centripetal force \cite%
{liu16,liu17-1}. One is familiar to us \cite{liu16,liu17-1},%
\begin{equation}
\frac{d}{dt}\mathbf{v}=-\mathbf{n}\frac{v^{2}}{R},  \label{1}
\end{equation}%
where $1/R$ stands for the first local curvature of the geodesic along which
the classical particle is bound to move, and $\mathbf{v}$ represents the
velocity of the particle, $\mathbf{n}$ symbolizes the normal vector on the
surface, and $t$ denotes the time as usual. To transition to quantum
mechanics, it is better to write it into the following form depending on
momentum $\mathbf{p}=\mu \mathbf{v}$ instead of $\mathbf{v}$, with\ $\mu $
being the mass of the particle and $H=p^{2}/(2\mu )=\mu v^{2}/2$ being the
Hamiltonian of the free particle, 
\begin{equation}
\frac{d}{dt}\mathbf{p}=-2\frac{\mathbf{n}}{R}H.  \label{2}
\end{equation}%
The main point of the present paper is to explore whether the Ehrenfest
theorem holds true. If true, we should have in quantum mechanics, 
\begin{equation}
\lbrack \mathbf{p,}H]=-i\hbar (\frac{\mathbf{n}}{R}H+H\frac{\mathbf{n}}{R}).
\label{3}
\end{equation}%
It is really true for some special cases, e.g., the spherical surface \cite%
{liu13-1}. However, whether it is true in general remains an open problem.
What is more, if one attempts to start from the equation (\ref{3}), the
first formidable problem is that the first curvature $1/R$ of the geodesic
in (\ref{3}) is hard to be attainable except for some simple cases such as
spherical surface, etc. Fortunately, this problem is not fatal. The equation
(\ref{3}) in fact combines both the intrinsic and extrinsic descriptions.
Not referring to dependence on the geodesic, the equation (\ref{1}) or (\ref%
{2}) becomes, in classical mechanics \cite{ikegami,weinberg},%
\begin{equation}
\frac{d}{dt}\mathbf{p=}-\mathbf{n}\left( \frac{\mathbf{\mathbf{p}}\cdot
\nabla \mathbf{\mathbf{n}}\cdot \mathbf{\mathbf{p}}}{\mu }\right) =-2\mathbf{%
n}S,  \label{4}
\end{equation}%
where, noting $n_{ji}=n_{ij}$ because, as we see shortly, we can always
choose a surface function $f(x)=0$ such that $\left\vert \nabla
f(x)\right\vert =1$ so we have $n_{i}=\left( \nabla f(x)\right)
_{i}=\partial _{i}f(x)$ and $n_{ij}\equiv \partial _{j}\partial _{i}f(x)$ ,%
\begin{equation}
S=\frac{\mathbf{\mathbf{p}}\cdot \nabla \mathbf{\mathbf{n}}\cdot \mathbf{%
\mathbf{p}}}{2\mu }=\frac{1}{2\mu }\mathbf{\mathbf{p}}\cdot \nabla \mathbf{%
\mathbf{n}}\cdot \mathbf{\mathbf{p}}=\frac{1}{2\mu }p_{i}n_{ij}p_{j}.
\label{5}
\end{equation}%
Thus, an examination of relationship between quantum version of (\ref{4}),
i.e., Eq. (\ref{11}), and Eq. (\ref{4}) itself suffices to see whether the
Ehrenfest theorem holds true. To achieve our goals, we perform our analysis
carefully.

First, we stress that a belief is widely hold which claims that the
Ehrenfest theorem is still valid. Noticing that the equation (\ref{4}) has
another apparently different form, used by, for instance, Weinberg \cite%
{weinberg}. Let us explain how it is. We assume the surface equation to be $%
f(x)=0$, where $f(x)$ is some smooth function of position $\mathbf{x\equiv }%
(x_{1},x_{2},...x_{N})$ in $R^{N}$, whose normal vector is $\mathbf{n}\equiv
\nabla f(x)/|\nabla f(x)|$. We have for (\ref{4}) and (\ref{5}),
respectively, 
\begin{equation}
\frac{d}{dt}\mathbf{p=-}2\frac{\nabla f(x)}{|\nabla f(x)|}S,  \label{6}
\end{equation}%
where%
\begin{eqnarray}
S &\mathbf{=}&\frac{1}{2\mu }\mathbf{\mathbf{p}}\cdot \nabla \mathbf{\mathbf{%
n}}\cdot \mathbf{\mathbf{p}}  \notag \\
&\mathbf{\mathbf{=}}&\frac{1}{2\mu }\frac{\mathbf{\mathbf{p}}\cdot \nabla
\nabla f(x)\cdot \mathbf{\mathbf{p}}}{|\nabla f(x)|}  \notag \\
&=&\frac{1}{2\mu }\frac{\mathbf{\mathbf{p}}\cdot \nabla \left( \mathbf{%
\mathbf{p}}\cdot \nabla f(x)\right) }{|\nabla f(x)|}  \notag \\
&=&\frac{1}{2\mu }\frac{\left( \mathbf{\mathbf{p}}\cdot \nabla \right)
^{2}f(x)}{|\nabla f(x)|}.  \label{7}
\end{eqnarray}%
So the expression taken by Weinberg \cite{weinberg} is reached,%
\begin{equation}
\frac{d}{dt}\mathbf{p=-}\frac{\nabla f(x)}{\mu \left( \nabla f(x)\right) ^{2}%
}\left( \mathbf{\mathbf{p}}\cdot \nabla \right) ^{2}f(x).  \label{8}
\end{equation}%
In classical mechanics two forms (\ref{4}) and (\ref{8}) are identical, but,
most importantly, in quantum mechanics these two forms are totally different
because of operator-orderings. Weinberg asserts without justification that
the quantum mechanics "\textit{yields the same equation of motion}" as
provided by classical mechanics \cite{weinberg}, 
\begin{equation}
\frac{d}{dt}\mathbf{p=}\frac{1}{i\hbar }[\mathbf{p,}H]=-\frac{\nabla f(x)}{%
\mu \left( \nabla f(x)\right) ^{2}}\left( \mathbf{\mathbf{p}}\cdot \nabla
\right) ^{2}f(x).  \label{9}
\end{equation}%
The equation of motions taking the same form in both quantum and classical
mechanics implies nothing but the Ehrenfest theorem. We argue that this
assertion is dubious. The operator of the right-handed side of this equation
is not manifestly hermitian once $\left\vert \nabla f(x)\right\vert \neq 1$, 
\begin{equation}
\left( \frac{\nabla f(x)}{\left( \nabla f(x)\right) ^{2}}\left( \mathbf{%
\mathbf{p}}\cdot \nabla \right) ^{2}f(x)\right) ^{\dag }\neq \frac{\nabla
f(x)}{\left( \nabla f(x)\right) ^{2}}\left( \mathbf{\mathbf{p}}\cdot \nabla
\right) ^{2}f(x).  \label{10}
\end{equation}%
There are too many uncontrollable operator-orderings. Among them, we like to
write a few, 
\begin{subequations}
\begin{eqnarray}
&&\frac{1}{2}\left( 
\begin{array}{c}
\frac{\nabla f(x)}{\left( \nabla f(x)\right) ^{2}}\left( \mathbf{\mathbf{p}}%
\cdot \nabla \right) ^{2}f(x) \\ 
+\left( \frac{\nabla f(x)}{\left( \nabla f(x)\right) ^{2}}\left( \mathbf{%
\mathbf{p}}\cdot \nabla \right) ^{2}f(x)\right) ^{\dag }%
\end{array}%
\right) ,  \label{oo1} \\
&&\frac{1}{2}\left( 
\begin{array}{c}
\nabla f(x)\left( \mathbf{\mathbf{p}}\frac{1}{\nabla f(x)}\cdot \nabla
\right) ^{2}f(x) \\ 
+\left( \nabla f(x)\left( \mathbf{\mathbf{p}}\frac{1}{\nabla f(x)}\cdot
\nabla \right) ^{2}f(x)\right) ^{\dag }%
\end{array}%
\right) ,  \label{oo2} \\
&&\frac{1}{2}\left( 
\begin{array}{c}
\left( \mathbf{\mathbf{p}}\frac{1}{\nabla f(x)}\cdot \nabla \right)
^{2}f(x)\nabla f(x) \\ 
+\left( \left( \mathbf{\mathbf{p}}\frac{1}{\nabla f(x)}\cdot \nabla \right)
^{2}f(x)\nabla f(x)\right) ^{\dag }%
\end{array}%
\right) ,etc.  \label{oo3}
\end{eqnarray}%
If the Ehrenfest theorem is true, we do not know what operator-orderings
and/or their combinations make sense \cite{liu17-1,damao1,damao2}.

Secondly, we stress that physics depends on the geometric invariants rather
than those changing along with the coordinate transformation or those
changing with the form of the surface equation chosen. No matter what form
of the surface equation we begin with, once there is a normal vector $%
\mathbf{n}$, only the unit normal vector and/or its derivatives enter the
physics equation. From the point of view of physics, we can always choose
the equation of the surface such that $|\nabla f(x)|=1$ \cite{ikegami,liu18}%
, so that $\mathbf{n}\equiv \nabla f(x)$. The form (\ref{4}) is therefore
advantageous over that (\ref{8}). In other words, on the grounds of physics
or mathematics, if some form of the (\ref{9}) turns out to be true, results
must be independent of peculiar form of $f(x)$. In quantum mechanics the
equation (\ref{4}) takes a simple form because of $S=S^{\dag }$, 
\end{subequations}
\begin{equation}
\frac{1}{i\hbar }[\mathbf{p,}H]=-\left( \mathbf{n}S+S\mathbf{n}\right) .
\label{11}
\end{equation}

Thirdly, the remaining problem is: whether this equation (\ref{11}) holds
true. We are going to show that with known $\mathbf{p}$ and $H$, this
equation breaks down. To note that there are independent approaches to give
the form of the momentum $\mathbf{p}$ and Hamiltonian $H$. The momentum $%
\mathbf{p}$ is the so-called the geometric one, \cite%
{liu11,liu13-1,liu13-2,waveguide,wang17}, 
\begin{equation}
{\mathbf{p}}=-i\hbar ({\nabla _{S}}+\frac{{M{\mathbf{n}}}}{2}),  \label{GM}
\end{equation}%
where ${\nabla _{S}}$ is the gradient operators in $R^{N}$ and $\Sigma ^{N-1}
$ respectively, and ${M\equiv }-\nabla {\cdot }$ ${{\mathbf{n}}}$ is the
mean curvature with ${\nabla }$ being the gradient operators in $R^{N}$. As
to Hamiltonian $H$, we have in general, \ 
\begin{equation}
H=-\frac{\hbar ^{2}}{2\mu }\nabla _{LB}^{2}+V_{G},
\end{equation}%
where $\nabla _{LB}^{2}={\nabla _{S}}\cdot {\nabla _{S}}$ is the
Laplace-Beltrami operator which is the dot product of the gradient operator $%
{\nabla _{S}\equiv }\nabla _{N}-\mathbf{n}\partial _{n}$ on the surface $%
\Sigma ^{N-1}$ and $V_{G}$ is curvature-induced potential dependent on both $%
{M}$ and $K{\equiv }\nabla {{\mathbf{n:}}}\nabla {{\mathbf{n}}}$, via two
real constants $\xi $ and $\eta $, 
\begin{equation}
V_{G}=-\frac{\hbar ^{2}}{8\mu }\left( \xi K-\eta {M}^{2}\right) .
\end{equation}%
Two real constants $\xi $ and $\eta $ depend on theoretical considerations.
What most of us are familiar with are two cases, 
\begin{equation}
H=-\frac{\hbar ^{2}}{2\mu }\nabla _{LB}^{2},\text{ and }H=\frac{p^{2}}{2\mu }%
,
\end{equation}%
corresponding to $\left( \xi ,\eta \right) =\left( 0,0\right) $ and $\left(
\xi ,\eta \right) =\left( 0,1\right) $, respectively. In fact, by the
so-called \textit{confining potential formalism }\cite{wang17,jk,dacosta,fc}%
\textit{, }we have $\left( \xi ,\eta \right) =\left( 2,1\right) $. By the
so-called \textit{abelian conversion} \cite{KS} for spherical surface, one
obtains$\left( \xi ,\eta \right) =\left( 0,0\right) $. Some simply makes
assumption\textit{\ }\cite{weinberg,HIM}, $\left( \xi ,\eta \right) =\left(
0,1\right) $.

In final, after lengthy computations, we have following result,%
\begin{equation}
\frac{1}{i\hbar }[\mathbf{p,}H]+\left( \mathbf{n}S+S\mathbf{n}\right) \equiv 
\mathbf{F}\neq 0,
\end{equation}%
where,%
\begin{eqnarray}
\mathbf{F} &=&-\frac{\hbar ^{2}}{4\mu }\left( \mathbf{n}\nabla ^{2}M\right. 
\notag \\
&&\mathbf{+(}2-\xi \mathbf{)}\left\{ \left( \nabla {{\mathbf{n:}}}\nabla
\nabla \right) {{\mathbf{n-}}}\nabla {{\mathbf{n:}}}\left( {{\mathbf{n}}}%
\cdot \nabla \nabla {{\mathbf{n}}}\right) \right\}   \notag \\
&&+\left. \mathbf{(}1-\eta \mathbf{)}M\left\{ \nabla ^{2}{{\mathbf{n-n}}}%
\cdot \nabla M\right\} \right) .
\end{eqnarray}%
It is in general not zero, and it is evident that among different
theoretical considerations, the simplest case is $\left( \xi ,\eta \right)
=\left( 2,1\right) $ \cite{liu17-2}. A curious question follows: Is there
any curvature-induced potential that makes $\mathbf{F}=0$? The answer is
negative. To see it, let us assume that there is such a potential $G$ in the
Hamiltonian, \ 
\begin{equation}
H=-\frac{\hbar ^{2}}{2\mu }\nabla _{LB}^{2}+G,
\end{equation}%
which renders following equation hold true no matter how the surface is
curved, 
\begin{equation}
\frac{1}{i\hbar }[\mathbf{p,}H]+\left( \mathbf{n}S+S\mathbf{n}\right) =0.
\end{equation}%
If it is true, the curvature-induced potential $G$ must consist of two
parts, in which one is $V_{G}$ with $\left( \xi ,\eta \right) =\left(
2,1\right) $, and rest is $W\equiv $ $G-V_{G}$. We have, however, 
\begin{equation}
\frac{1}{i\hbar }[\mathbf{p,}H]+\left( \mathbf{n}S+S\mathbf{n}\right) =-%
\frac{\hbar ^{2}}{4\mu }\mathbf{n}\nabla ^{2}M-\left\{ \nabla W-\mathbf{n}%
\left( \mathbf{n\cdot }\nabla W\right) \right\} ).
\end{equation}%
These two terms in right-handed side are orthogonal to each other. To be
precise, the first term is along the normal direction but the second term is
on the tangential surface for $\mathbf{n\cdot }\left\{ \nabla W-\mathbf{n}%
\left( \mathbf{n\cdot }\nabla W\right) \right\} =0$. The result is contrary
to the assumption. Thus, we see that the Ehrenfest theorem breaks down.

In conclusion, in quantum mechanics for the particle on the hypersurface
there is certainly some forms of the curvature driven force. In contrast to
the widely hold belief, the Ehrenfest theorem cannot be applied to the
particle on the curved space.

\begin{acknowledgments}
This work is financially supported by National Natural Science Foundation of
China under Grant No. 11675051.
\end{acknowledgments}

\end{document}